# XrdML, a new way to store (and exchange) X-ray powder diffraction measurement data

Transparent and open – these keywords describe the new approach that we (PANalytical) adopt with our new data platform for the X'Pert XRD software. Based on the universal format for structured documents and data XML (eXtended Markup Language), this approach will give any user complete control over his XRD measurement data.

Traditionally, XRD measurement data were stored in binary files. Although this saved valuable disk space, it allowed only limited possibilities for analysis. In 1997 PANalytical introduced the current X'Pert software, which stored the measurement data in a database. This allows for a wide range of pre-programmed features for analysis and processing. With the introduction of the XRDML data platform, PANalytical takes this development a step further, giving users complete access to their measurement results.

In the new XRDML data platform, ASCII format XRD measurement data are stored in XML based files (.XRDML files) that contain all measurement data as well as all information required to reproduce the data, including the instrument type and settings.

The data are written in accordance with the XML schema, which is described at the website www.xrdml.com upon release of the package. The information contained in the schema gives users the possibility to develop tailor-made programs to process their measurement data in any way they need. The new software platform allows users of XRD systems to share their measurements with others, such as colleagues, fellow researchers, or authorities. All they need is a standard Internet browser to view the measurement data. Sharing XRD results has never been easier; the user is not limited by proprietary formats. Since the applicable XML schema includes all information about the measurement and the equipment used, the XRDML platform allows complete trace ability of results. This trace ability of results, in principle, is an important condition for compliance with FDA Part 11, which is particularly important for the pharmaceutical industry The XrdML approach also allows for a seamless integration with the Windows Explorer (with tool tip information, viewing, conversion and reporting tools and thumbnail views of the diffractograms!). Furthermore the user can create own style sheets to generate reports just the way he wants them.

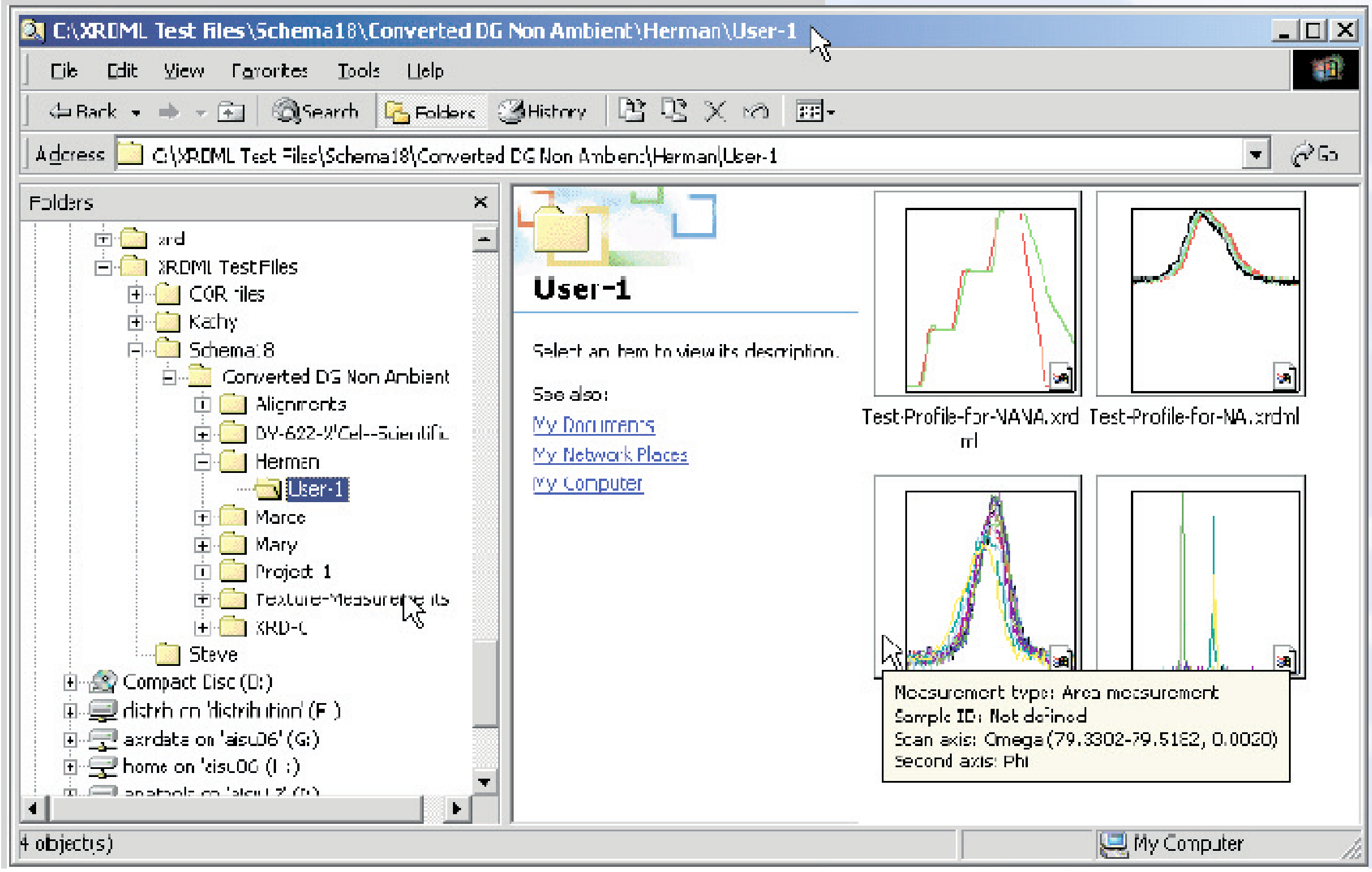

*Example of Thumbnail view of XRD measurements in the Windows Explorer.*

## About XML

XML (Extensible Markup Language) is a universal format for structured documents and data, somewhat similar to HTML. It is an industry standard controlled by the World Wide Web Consortium (www.w3c.org), and supported by well-known software makers with a wide and growing range of software tools and utilities.

XML uses 'tags' and 'attributes' to delimit and describe pieces of data in an ASCII readable text file. XML 'schemas' expresses shared vocabularies and allow machines to carry out rules made by people. They provide a means for defining and validating the structure, content and semantics of XML documents. XML is license-free, and allows users to build their own application. The large and growing industry support means that users are never tied to a single vendor.

# *Layout of the XrdML Measurement Schema (shown in XML Spy ©)*

## *1) Root level view:*

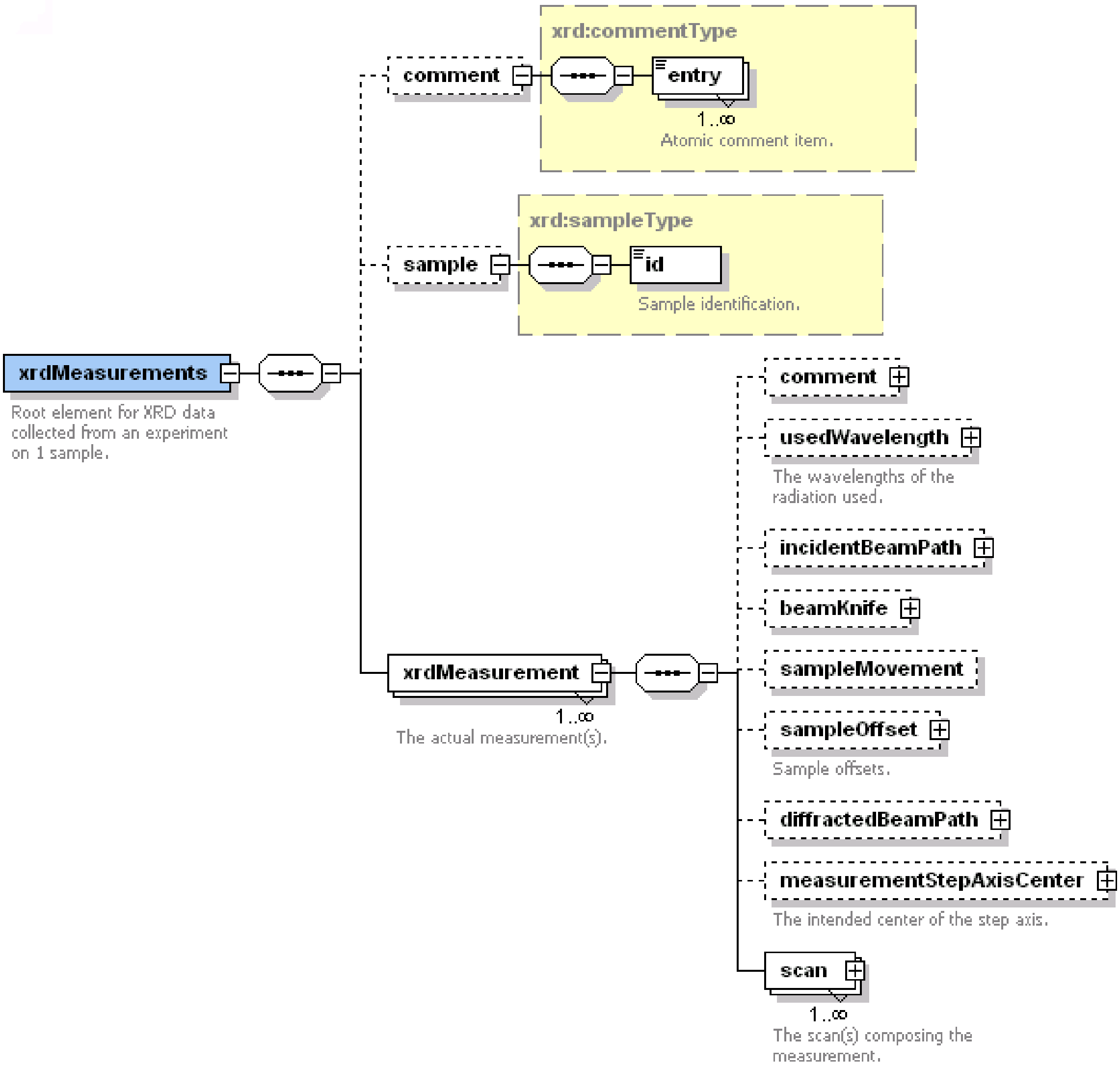

## *2) Detailed design of Incident Beam Path and X-ray Mirror:*

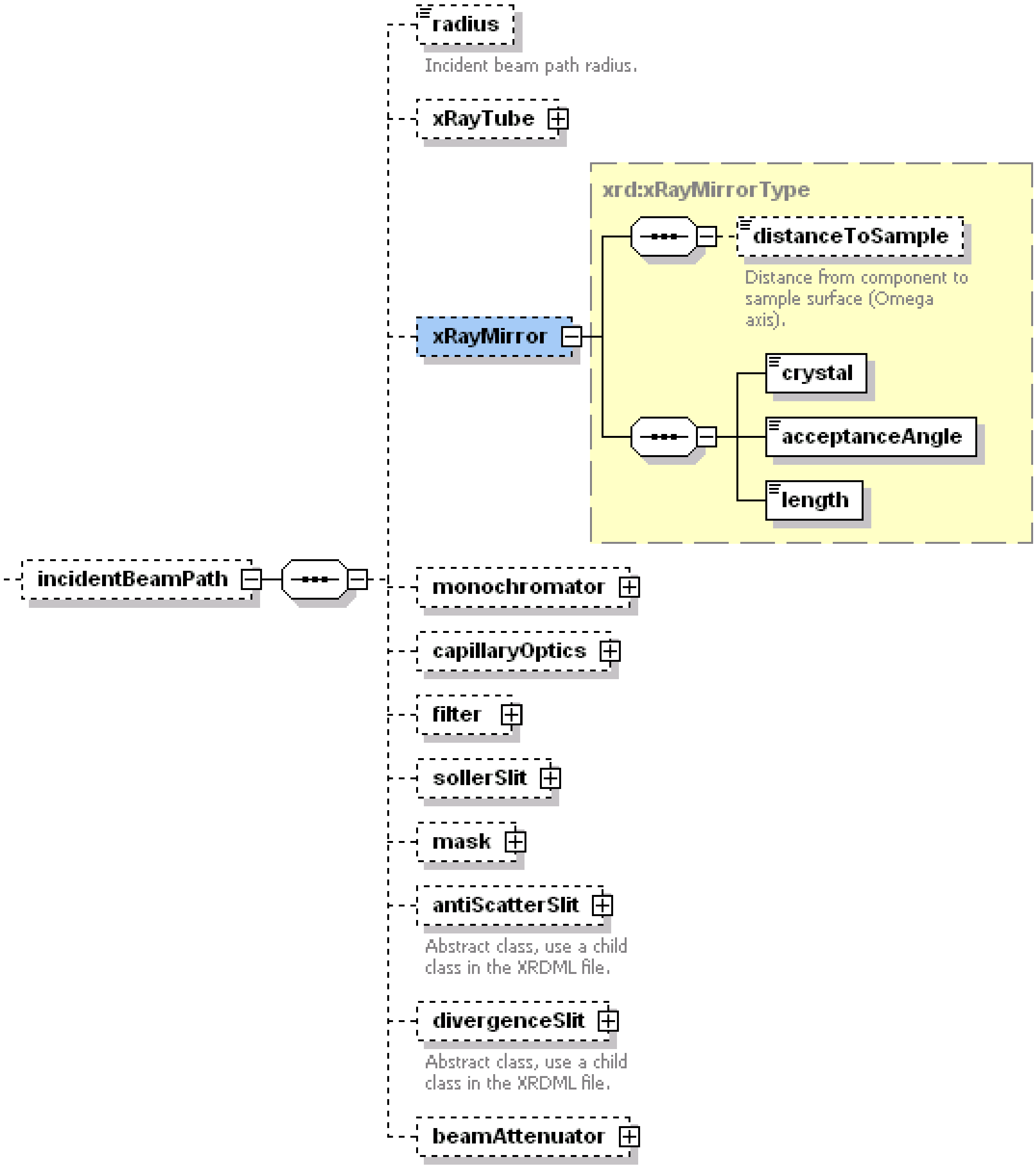

### *3) Detailed design of Diffracted Beam Path and Detector:*

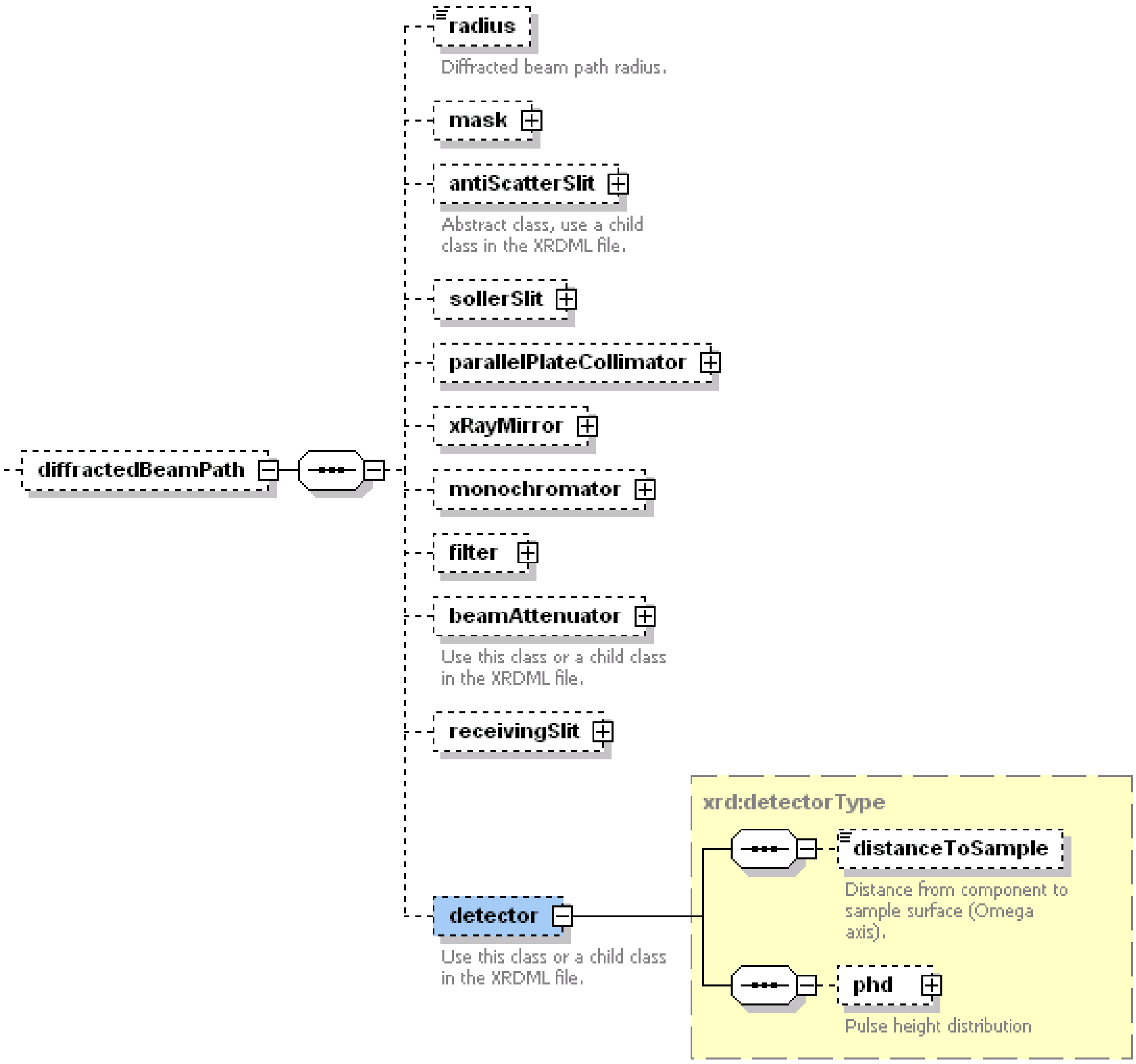

## *4) Detailed design of Scan and Non Ambient Data Points:*

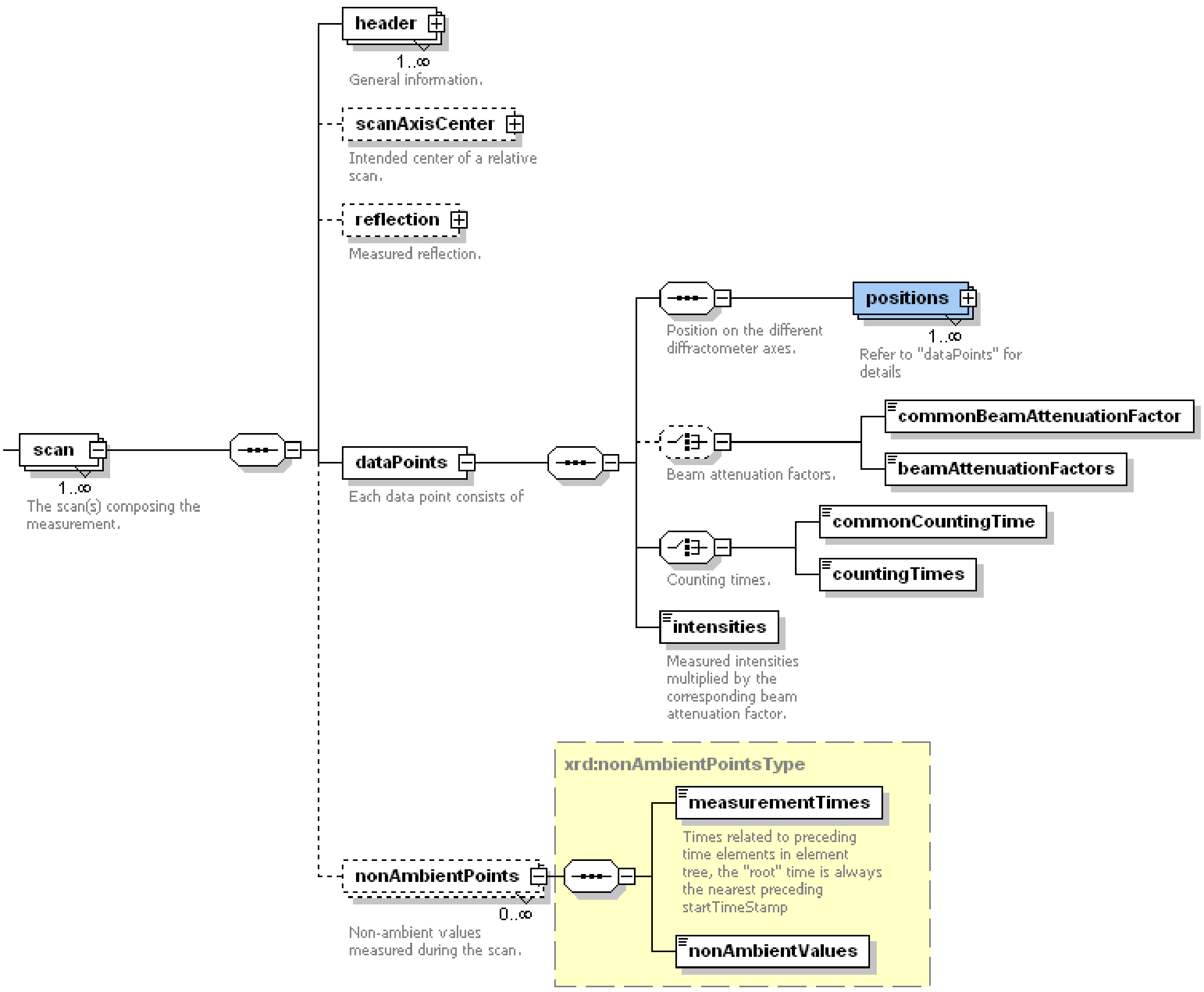

## *5) Detailed design of ordinary Data Points:*

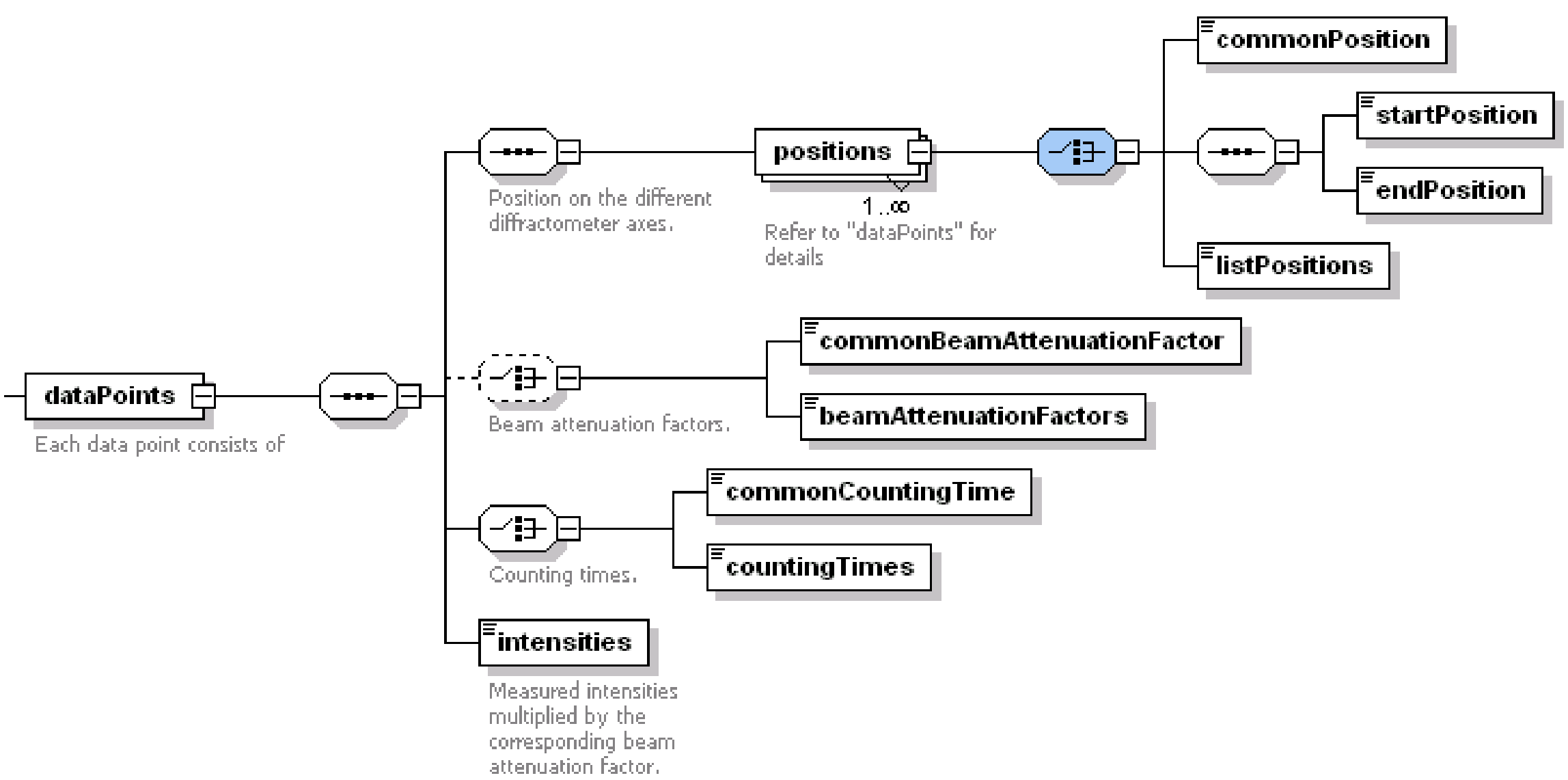

# *Example of an XrdML file (shown in XML Spy ©)*

## *1) Overview of complete Measurement:*

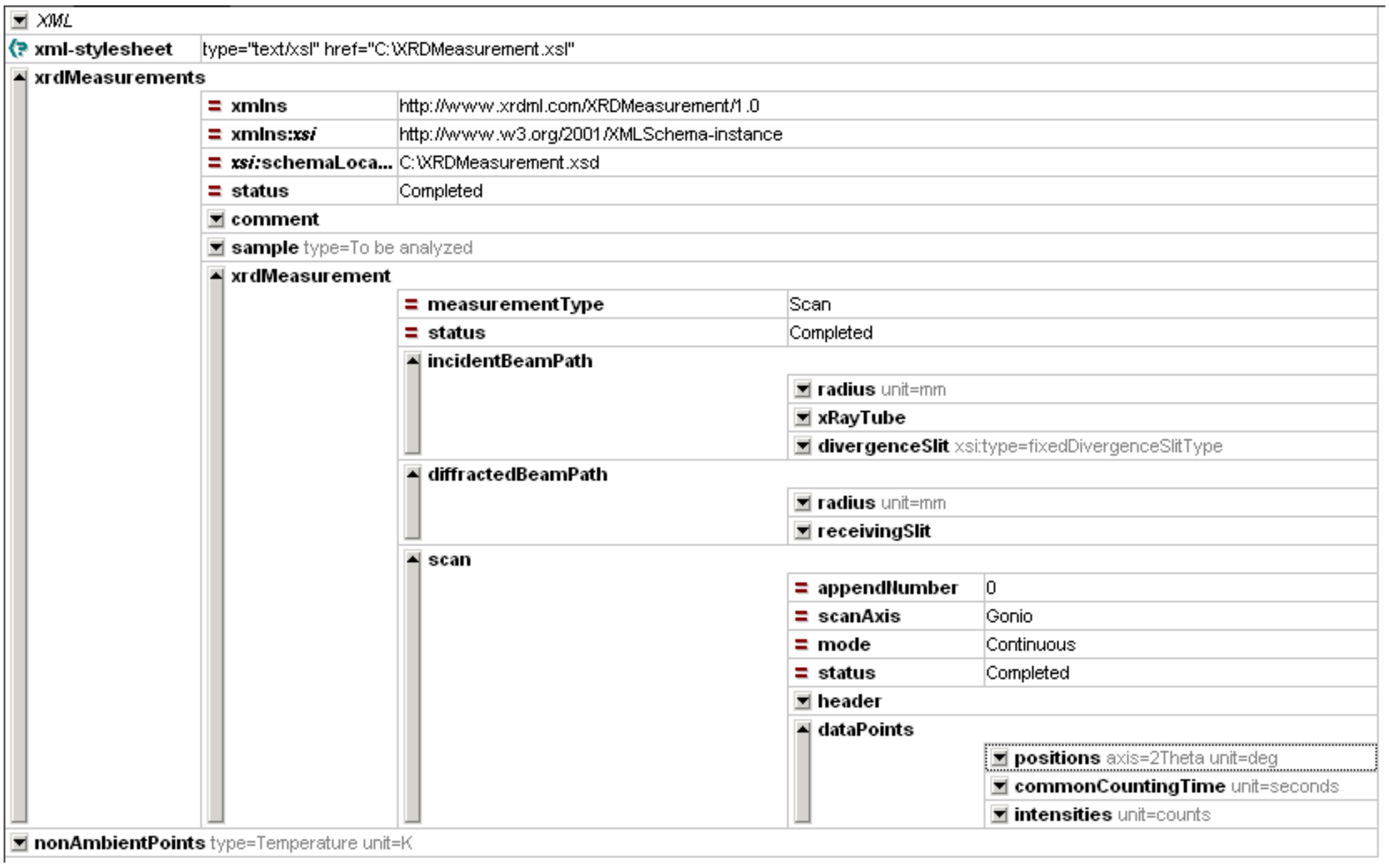

## *2) Detailed view of Scan and Data Points:*

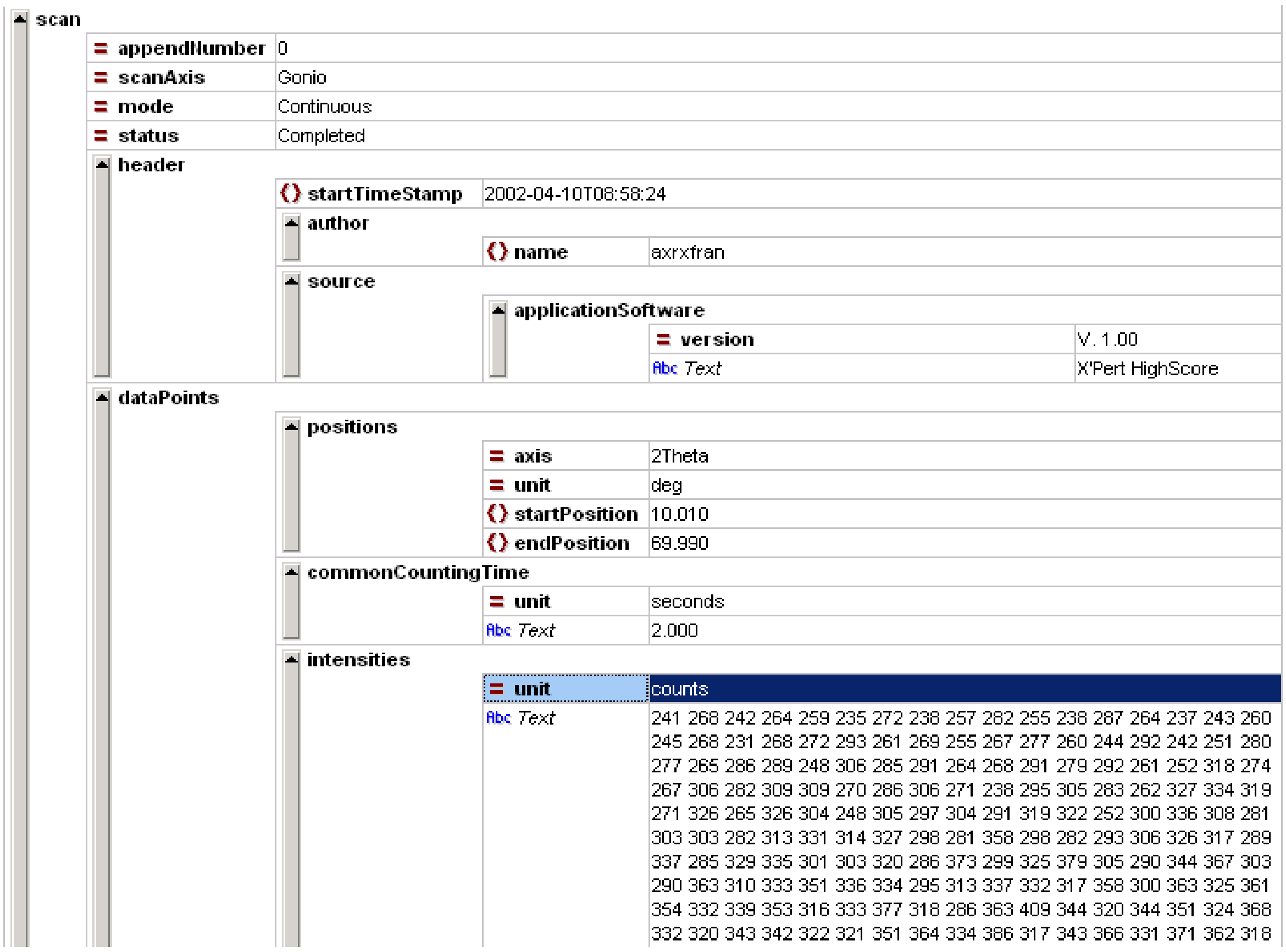